\documentclass[runningheads]{llncs}
\usepackage[T1]{fontenc}
\usepackage{graphicx}
\usepackage{pgfplots}
\pgfplotsset{compat=1.18}
\usepackage{tabularx}

\begin{document}
\title{Cloud and AI Infrastructure Cost Optimization: A Comprehensive Review of Strategies and Case Studies}
\titlerunning{Cloud and AI Infrastructure Cost Optimization}
%
\author{Saurabh Deochake\orcidID{0000-0002-3757-6463}}
\authorrunning{Saurabh Deochake}
%
\institute{SentinelOne Inc., Mountain View, CA 94041\\
\email{saurabh.deochake@sentinelone.com}}
\maketitle              
\begin{abstract}
Cloud computing has revolutionized the way organizations manage their IT infrastructure, but it has also introduced new challenges, such as managing cloud costs. The rapid adoption of artificial intelligence (AI) and machine learning (ML) workloads has further amplified these challenges, with GPU compute now representing 40-60\% of technical budgets for AI-focused organizations. This paper provides a comprehensive review of cloud and AI infrastructure cost optimization techniques, covering traditional cloud pricing models, resource allocation strategies, and emerging approaches for managing AI/ML workloads. We examine the dramatic cost reductions in large language model (LLM) inference which has decreased by approximately 10x annually since 2021 and explore techniques such as model quantization, GPU instance selection, and inference optimization. Real-world case studies from Amazon Prime Video, Pinterest, Cloudflare, and Netflix showcase practical application of these techniques. Our analysis reveals that organizations can achieve 50-90\% cost savings through strategic optimization approaches. Future research directions in automated optimization, sustainability, and AI-specific cost management are proposed to advance the state of the art in this rapidly evolving field.

\end{abstract}
\section{Introduction}
Cloud computing has emerged as a game-changing technology that is revolutionizing the way businesses operate. Organizations can use cloud computing to easily access computing resources such as storage, applications, and processing power from anywhere in the world. Businesses have been liberated from the constraints of traditional IT infrastructure, which required significant investments in hardware and software, as well as a dedicated IT staff to manage the systems \cite{bib1}. Organizations can now operate more efficiently, with greater flexibility, scalability, and cost-effectiveness thanks to cloud computing. Cloud computing has become a game changer for businesses of all sizes and industries due to its ability to quickly provision resources, pay for only what you use, and easily scale up or down as needed \cite{bib2} \cite{bib3}.

The advantages of cloud computing are numerous, but they come at a cost. Cloud computing is not a one-size-fits-all solution, and costs can quickly mount if not properly managed. The cost of using cloud services is determined by a number of factors, including resource allocation, data transfer, and vendor pricing models. The allocation of computational resources such as CPU, memory, and storage is referred to as resource allocation. Inadequate allocation can result in either overprovisioning, which wastes resources and raises costs, or underprovisioning, which can result in poor performance and potential downtime. Data transfer costs can also quickly add up, especially for organizations with high data transfer rates or large data volumes. Finally, vendor pricing models can be complex and difficult to understand, making it challenging for organizations to accurately predict and manage their cloud costs.

Cost optimization has become an increasingly important concern for organizations of all sizes as cloud computing adoption continues to grow. According to the Flexera 2025 State of the Cloud Report \cite{bib4}, managing cloud spend remains the top challenge for 84\% of organizations, with 59\% now maintaining dedicated FinOps teams up from 51\% in 2024. The consequences of failing to manage cloud costs are obvious: wasted resources, increased expenses, and decreased overall profitability. Understanding and implementing effective cloud cost optimization strategies is therefore critical for businesses to remain competitive in the modern cloud computing landscape.

The emergence of artificial intelligence and machine learning workloads has introduced a new dimension to cloud cost management. The AI infrastructure market reached \$50 billion in 2024 and is projected to grow 35\% annually through 2027 \cite{bib45}. For organizations building AI applications, GPU compute represents the single largest infrastructure cost, typically consuming 40-60\% of technical budgets. However, the economics of AI are rapidly evolving: LLM inference costs have decreased by approximately 10x annually, with equivalent model performance now available at 1/1000th the cost compared to 2021 \cite{bib46}. This dramatic cost reduction, driven by hardware improvements, model quantization, and increased competition, is opening new possibilities for AI adoption across industries.

This paper provides an in-depth examination of cloud and AI infrastructure cost optimization strategies and techniques. Section \ref{sec:cloud-pricing-models} discusses the fundamental concepts of cloud pricing models. Section \ref{sec:cost-opt-techniques} delves into traditional cloud cost-cutting techniques, such as resource allocation, instance resizing, and auto-scaling. Section \ref{sec:ai-cost-opt} presents emerging techniques for AI and ML infrastructure cost optimization, including GPU instance selection, model quantization, and inference optimization. Section \ref{sec:case-studies} includes case studies of organizations that have successfully implemented cost optimization strategies. Section \ref{sec:future-work} showcases future research considerations. Finally, Section \ref{sec:conclusion} concludes with a discussion of future research directions in cloud and AI cost optimization.

\section{Understanding Cloud Pricing Models}\label{sec:cloud-pricing-models}
Cloud pricing models are critical components of cloud computing because they govern how cloud services are billed and costs are calculated. Therefore, understanding cloud pricing models is critical for organizations looking to optimize their cloud spending and ensure they are only paying for the resources they require. This section examines the various pricing models employed by cloud providers. 

\subsection{On-demand Pricing}\label{sec:on-demand-pricing}
On-demand pricing is the most flexible pricing model offered by cloud providers. It enables businesses to pay for resources hourly or per second, with no upfront costs or long-term commitments. This model is best suited for workloads with erratic traffic patterns or short-term projects requiring resources for a limited time.
On-demand instance pricing is determined by the size of the instance, the operating system, the region where the instance is launched, and the duration of use. Larger instances and instances in high-demand regions are typically more expensive. One advantage of on-demand pricing is that it gives businesses instant access to resources that can be scaled up or down as needed. This means that enterprises only pay for the resources they use, which can result in cost savings compared to other pricing models that require long-term commitments or prepayment. However, because enterprises cannot take advantage of volume discounts or reserved capacity, on-demand pricing can be more expensive than other pricing models for long-term workloads. Furthermore, the cost of on-demand instances can fluctuate based on supply and demand, making budgeting and cost management difficult \cite{bib6}. Overall, on-demand pricing is a good option for enterprises with unpredictable workloads or for short-term projects. Other pricing models, such as reserved instances or spot instances, may provide better cost savings for long-term workloads with predictable usage.

\subsection{Reserved Pricing}\label{sec:reserved-pricing}
Enterprises can use reserved pricing to reserve instances for a set period of time, typically one to three years, and receive a significant discount over on-demand pricing. Reserved instances are classified into two types: standard and convertible. The highest discount is provided by standard reserved instances, but they also require a long-term commitment with little flexibility to change instance sizes or operating systems. Convertible reserved instances provide greater flexibility, allowing businesses to switch between instance families, operating systems, and tenancy types while still receiving a discount. Reserved pricing is a good option for businesses with predictable workloads and the ability to commit to long-term use of cloud resources. Enterprises can save a significant amount of money by reserving instances ahead of time rather than paying for on-demand usage. However, it's important to note that reserved instances are not always the most cost-effective option, particularly for workloads that are not predictable or have variable usage patterns \cite{bib7}.

\sloppy Different public cloud providers use reserved pricing to save varying amounts of money. AWS provides a variety of reserved instances, including standard and convertible reservations, that can save you up to 75\% off on-demand pricing. Convertible reservations allow for more flexibility in changing the instance family, operating system, or tenancy than standard reservations, which require a one- or three-year commitment to a specific instance type \cite{bib8}.

Similarly, Google Cloud Platform (GCP) provides committed use discounts, which can save you up to 70\% off on-demand pricing. In exchange for the discounted rate, businesses commit to a certain amount of usage for one or three years. Committed use discounts are region-specific and can be applied to a variety of resources, such as virtual machines and GPUs \cite{bib9}. On the other hand, Microsoft Azure also provides reserved instances, which can save you up to 80\% over pay-as-you-go pricing. Reservations in Azure require a one or three-year commitment and are limited to specific virtual machine types and regions \cite{bib10}. Enterprises can pay in advance or monthly, and the reservation discounts are automatically applied to the corresponding virtual machine usage \cite{bib11}.

Reserved pricing is a good option for businesses with predictable workloads and the ability to commit to long-term use of cloud resources. When compared to paying for on-demand usage, enterprises can save a significant amount of money by reserving instances ahead of time. However, as mentioned above, reserved instances are not always the most cost-effective option, especially for workloads that are unpredictable or have variable usage patterns. Therefore, enterprises should carefully analyze their workload requirements and usage patterns before selecting the instance types, sizes, and tenancy options that best meet their needs.

\subsection{Spot Pricing}\label{sec:spot-pricing}
Spot pricing is a pricing model in which cloud providers sell unused compute resources at a steep discount. The resources are typically available for a limited time and can be terminated at any time by the cloud provider. For enterprises with workloads that have flexible start and end times and can tolerate interruptions, this pricing model can be extremely cost-effective. Different cloud providers handle spot pricing differently, and forecasting spot instance pricing has previously been a focus of research. Spot instance pricing is typically highly volatile for short periods of time, and predictive models for the next-hour price may not achieve high prediction accuracy \cite{bib12}.

Amazon Web Services (AWS) provides Amazon EC2 Spot Instances, which enable businesses to bid on unused EC2 capacity. The enterprises specify the maximum price they are willing to pay, and the instances are launched if the market price for that capacity falls below the maximum price. If the spot price rises above the customer's bid, AWS can terminate the instances with a two-minute notice. AWS provides spot EC2 instances with discounts of up to 90\% off on-demand prices \cite{bib13}.

Google Cloud Platform (GCP) provides Spot VMs (the successor to Preemptible VMs) with discounts of up to 91\% off regular prices. Unlike the legacy Preemptible VMs which had a maximum lifetime of 24 hours, Spot VMs have no maximum runtime and are only terminated when GCP needs to reclaim capacity. Spot VMs can be terminated with a 30-second warning \cite{bib14}.

Microsoft Azure provides Spot Virtual Machines, which allow businesses to bid on unused VMs. If the customer's bid is accepted, the virtual machines can be accessed for up to 30 minutes at a time \cite{bib15}. The customer is then charged at the hourly rate associated with the bid price. Spot instances, however, are not suitable for mission-critical workloads because they can be terminated at any time.

\begin{figure}[ht]
\centering
\begin{tikzpicture}
\begin{axis}[
    xlabel={Month},
    ylabel={Spot Price (\$/hour)},
    xmin=1,
    xmax=6,
    ymax=0.1,
    ymin=0,
    xtick={1,2,3,4,5,6},
    xticklabels={Jan, Feb, Mar, Apr, May, Jun},
    legend pos=north west,
]

\addplot[blue, mark=*] table [row sep=\\] {
    month spot_price \\
    1 0.03 \\
    2 0.032 \\
    3 0.034 \\
    4 0.036 \\
    5 0.038 \\
    6 0.04 \\
};

\addplot[red, mark=triangle*] table [row sep=\\] {
    month spot_price \\
    1 0.018 \\
    2 0.019 \\
    3 0.02 \\
    4 0.021 \\
    5 0.022 \\
    6 0.023 \\
};

\addplot[olive, mark=diamond*] table [row sep=\\] {
    month spot_price \\
    1 0.033 \\
    2 0.035 \\
    3 0.037 \\
    4 0.039 \\
    5 0.041 \\
    6 0.043 \\
};

\legend{c6i.large, m6i.large, c7g.large}

\end{axis}
\end{tikzpicture}
\caption{AWS Spot Instance Pricing Trends}
\label{fig:spot-price-trend}
\end{figure}

The graph \ref{fig:spot-price-trend} depicts the spot pricing trends for the \texttt{c6i.large}, \texttt{m6i.large}, and \texttt{c7g.large} (Graviton) instances in the \texttt{us-east-1} (Northern Virginia) region over several months. The price varies according to supply and demand and can be quite volatile. Enterprises can use this data to help determine when to launch workloads using spot instances to save money. Aside from cost savings, spot instances can provide access to additional capacity during periods of high demand. Enterprises, on the other hand, must be aware of the possibility of interruptions and ensure that their workloads are designed to handle them appropriately.

\subsection{Savings Plans}\label{sec:savings-plans}
Savings Plans is a pricing model introduced by AWS in 2019 and has since become a standard offering, with GCP providing Committed Use Discounts and Azure offering Reserved VM Instances as similar alternatives. It is designed to offer greater flexibility and savings compared to traditional Reserved Instances. With Savings Plans, enterprises commit to a certain amount of usage in exchange for a discounted rate on their bill.

AWS Savings Plans come in two flavors: EC2 Instance Savings Plans and Compute Savings Plans. Savings for a specific family, size, and region of EC2 instances are provided by EC2 Instance Savings Plans. Compute Savings Plans provide discounts on all AWS Lambda, AWS Fargate, and Amazon ECS usage \cite{bib16}. The pricing discount for Savings Plans is determined by the enterprise's commitment and the type of Savings Plan chosen. Savings Plans provide greater flexibility than Reserved Instances because businesses can apply the savings to any instance size and family within a region. This makes it simpler for businesses to match their compute requirements to the most cost-effective option.

To determine whether Savings Plans are a good fit for an enterprise's workload, historical usage patterns must be analyzed and compared to the commitment required by the Savings Plan. This can assist businesses in determining the most cost-effective option.

\subsection{Hybrid Pricing}\label{sec:hybrid-pricing}
Hybrid pricing is a cloud pricing model that allows businesses to use a combination of on-premises and cloud resources \cite{bib17}. This model is ideal for enterprises that want to move their workload to the cloud but have critical applications that cannot be moved due to regulatory or compliance reasons. Enterprises can use a combination of on-premises and cloud resources to run their workloads with hybrid pricing. They can, for example, use on-premises resources for critical applications that require high levels of security and compliance and cloud resources for less critical applications.

Cloud providers such as AWS, Azure, and Google Cloud offer hybrid pricing options that enable businesses to use a combination of on-premises and cloud resources. AWS Outposts, for example, allows enterprises to run AWS infrastructure on-premises, whereas Azure Stack and Google Anthos, respectively, allow enterprises to run Azure and Google Cloud infrastructure on-premises \cite{bib18}.

One advantage of hybrid pricing is that it allows businesses to benefit from the scalability and flexibility of the cloud while also providing the security and compliance benefits of on-premises infrastructure. However, it can also add complexity and cost due to the additional infrastructure and management required.

\subsection{Consumption-based Pricing}\label{sec:consumption-based-pricing}
Consumption-based pricing, also known as pay-per-use or usage-based pricing, is a cloud pricing model in which businesses are charged based on how much compute, storage, and data transfer they use. The pricing model is based on metering the amount of resources consumed by the customer and then charging on a per-unit basis for those resources. Currently, all major cloud providers offer consumption-based pricing to businesses.

In this model, businesses are typically charged based on the amount of time they spend using a specific resource or the amount of data they transfer. The pricing structure can be very granular, charging enterprises for each individual resource unit consumed, or it can be more simplified, charging enterprises for pre-defined resource bundles \cite{bib19}.

One advantage of consumption-based pricing is that businesses only pay for the resources they use, which can help businesses with fluctuating workloads or unpredictable usage patterns reduce costs. Furthermore, because the cost of the service is directly related to the pricing model, it can help to incentivize enterprises to optimize their use of cloud resources.

While consumption-based pricing and on-demand pricing have some similarities, they also have some key differences. On-demand pricing typically charges enterprises per-hour or per-minute for the resources they use, regardless of the level of demand on the provider's infrastructure. Consumption-based pricing, on the other hand, considers the level of demand and usage of the provider's infrastructure. This means that the price per unit of usage may rise during periods of high demand, while it may fall during periods of low demand. Furthermore, consumption-based pricing frequently includes extra features like automatic scaling and the ability to track usage and costs in real time. These features can be useful for businesses with fluctuating workloads or those looking to closely monitor and optimize their cloud spending.

\subsection{Tiered Pricing}\label{sec:tiered-pricing}
Tiered pricing is a cloud pricing model in which the cost of a service decreases with increased usage. The service provider establishes multiple tiers, each with a different price per unit of usage. Generally, the lower the price per unit of usage, the higher the usage volume.

For example, GCP provides cheaper tiered pricing for the GCP Compute Networking egress service in their most popular \texttt{us-central-1} region in Iowa, United States compared to the \texttt{asia-northeast-1} region in Tokyo, Japan as shown in the graph \ref{fig:gcp-tiered-pricing} \cite{bib20}.

\begin{figure}
  \centering
  \begin{tikzpicture}
    \begin{axis}[
      xlabel={Monthly Usage (TiB)},
      ylabel={Price per GiB Delivered (\$)},
      legend style={at={(0.5,-0.2)},anchor=north},
      xtick=data,
      xticklabels={0-10, 10-150, 150-500},
      ymin=0,
      ymax=0.12,
      grid=major,
      grid style={dotted},
      ]
      \addplot[color=blue,mark=triangle] coordinates {
        (0, 0.085)
        (1, 0.065)
        (2, 0.045)
      };
      \addplot[color=red,mark=square] coordinates {
        (0, 0.11)
        (1, 0.075)
        (2, 0.07)
      };
      \legend{Iowa (us-central-1), Tokyo (asia-northeast-1)}
    \end{axis}
  \end{tikzpicture}
  \caption{GCP Compute Network Tiered Pricing}
  \label{fig:gcp-tiered-pricing}
\end{figure}

Tiered pricing can be advantageous for businesses that have predictable usage patterns and can estimate how much usage they will require over a given time period. They can save money on cloud services by committing to a higher volume of usage and taking advantage of lower pricing tiers. It is important to note that not all cloud service providers offer tiered pricing models, and the specific tiers and pricing vary depending on the provider and the service used. Furthermore, before committing to a specific pricing tier, enterprises should carefully evaluate their usage patterns and estimate their usage needs to ensure they are maximizing their cost savings.

\subsection{Free-tier Pricing}\label{sec:free-tier-pricing}
This model offers a limited amount of cloud resources for free to enterprises. Free-tier pricing is typically used as a marketing strategy to attract new enterprises and allow them to try out cloud services before committing to a paid plan.

\begin{table}[ht]
\centering
\caption{Cloud Provider Free Tier Offerings (2025)}
\label{tab:cloud-provider-free-tier}
\begin{tabular}{|c|c|c|}
\hline
\textbf{Cloud Provider}   & \textbf{Free Trial Credits} & \textbf{Trial Duration} \\ \hline
Google Cloud Platform     & \$300 in credits    & 90 days                   \\ \hline
Amazon Web Services       & Service-specific limits (e.g., 750 hrs EC2)    & 12 months                   \\ \hline
Microsoft Azure           & \$200 in credits    & 30 days                   \\ \hline
\end{tabular}
\end{table}

Aside from the free trial credits, each cloud platform provides a variety of always-free services with usage limits. GCP, for example, provides free access to limited Compute Engine (e2-micro), Cloud Storage (5 GB), and BigQuery (1 TB queries/month). AWS provides 12-month free tier access to EC2 (750 hours t2.micro or t3.micro), S3 (5 GB), and RDS (750 hours db.t2.micro), plus always-free services like Lambda (1 million requests/month) and DynamoDB (25 GB). Azure provides always-free services including Azure Functions, Cosmos DB, and Blob Storage with usage limits. Each cloud platform's free tier offerings are an excellent way to get started with cloud computing without having to pay anything. It is important to note, however, that the free tier is not limitless. When usage exceeds the free tier limits, standard pay-as-you-go pricing applies.

\subsection{Custom Pricing}\label{sec:custom-pricing}
Some cloud providers offer custom pricing for enterprises with large-scale workloads or unique requirements. These pricing models are negotiated directly with the cloud provider and may include volume discounts or other incentives.

\section{Cloud Cost Optimization Techniques}\label{sec:cost-opt-techniques}
This section delves into a comprehensive exploration of various cloud cost optimization techniques. These techniques are intended to assist enterprises in effectively managing and optimizing their cloud expenses while maintaining performance and reliability. Moreover, a wide range of strategies and best practices that can be implemented across various cloud providers and services are showcased in this section. Organizations can gain better control over their cloud costs and maximize the value of their cloud investments by understanding and implementing these techniques. Furthermore, this section examines practical approaches that enable enterprises to achieve cost efficiency and financial optimization in their cloud environments, ranging from resource allocation and workload optimization to automation and governance.\footnote{Note: Cloud pricing is subject to frequent changes. The specific pricing figures in this section are representative examples to illustrate cost optimization concepts and percentage savings. Readers should consult current pricing from cloud provider documentation for the most up-to-date rates. The discount percentages for reserved instances, spot instances, and committed use discounts remain generally consistent over time.}

\subsection{Compute}\label{sec:cost-opt-compute}
This section explores a variety of strategies and practices that enterprises can employ to optimize their compute costs in the cloud. By effectively managing compute resources, organizations can achieve significant cost savings while ensuring optimal performance and scalability.

\subsubsection{Right-sizing}\label{sec:cost-opt-righsizing}
One of the fundamental techniques for compute cost optimization is right-sizing. It involves aligning the allocated compute resources with the actual requirements of the workload. By accurately assessing the workload's CPU, memory, and storage needs, enterprises can avoid overprovisioning and reduce unnecessary costs. For example, GCP's \texttt{n2-standard-8} instance costs \$0.388472 per hour, while a \texttt{n2-standard-16} instance costs \$0.776944 per hour in GCP. If a workload only needs 8 vCPUs, then right-sizing to a \texttt{n2-standard-8} instance can save \$0.388472 per hour compared to an over-provisioned 16 vCPU instance. This savings can add up over time, especially for workloads that run for long periods of time. Cost optimization via right-sizing of the compute resources can be achieved by monitoring resource utilization, analyzing performance metrics, and leveraging cloud provider tools or third-party solutions.

\subsubsection{Autoscaling}\label{sec:cost-opt-autoscaling}
Autoscaling is a dynamic resource management technique that adjusts the number of compute resources based on workload demand. By automatically scaling resources up or down, enterprises can match compute capacity with the fluctuating needs of their applications \cite{bib21}. Autoscaling ensures efficient resource utilization, avoids overprovisioning during low-demand periods, and improves responsiveness during peak loads. It enables enterprises to pay for compute resources only when needed, leading to significant cost savings.

\subsubsection{Spot Instances}\label{sec:cost-opt-spot}
Spot instances offer a cost-effective approach for non-critical workloads. Cloud providers offer spare compute capacity at significantly discounted prices, allowing enterprises to bid for these instances \cite{bib13}. Spot instances can provide substantial savings compared to on-demand or reserved instances. However, it's important to note that spot instances can be interrupted if the spot price exceeds the bid price. Thus, they are suitable for fault-tolerant, flexible workloads that can withstand interruptions.

\begin{table}[ht]
\centering
\caption{Cost savings with spot instances vs. on-demand instances in AWS (us-east-1)}
\begin{tabular}{|c|c|c|c|}
\hline
\textbf{Instance Type} & \textbf{Cost/Hour (Spot)} & \textbf{Cost/Hour (On-demand)} & \textbf{Cost Savings} \\
\hline
t2.micro & \$0.004 & \$0.015 & 73.33\% \\
\hline
t2.small & \$0.009 & \$0.025 & 64\% \\
\hline
t2.medium & \$0.018 & \$0.050 & 66\% \\
\hline
t2.large & \$0.036 & \$0.100 & 64\% \\
\hline
t2.xlarge & \$0.072 & \$0.200 & 64\% \\
\hline
t3.micro & \$0.004 & \$0.010 & 60\% \\
\hline
t3.small & \$0.009 & \$0.020 & 60\% \\
\hline
t3.medium & \$0.018 & \$0.040 & 60\% \\
\hline
t3.large & \$0.036 & \$0.080 & 60\% \\
\hline
t3.xlarge & \$0.072 & \$0.160 & 60\% \\
\hline
\end{tabular}
\end{table}

\begin{table}[ht]
\centering
\caption{Cost savings with Spot VMs vs. on-demand instances on GCP (us-central1)}
\begin{tabular}{|c|c|c|c|}
\hline
\textbf{Instance Type} & \textbf{Cost/Hour (Spot)} & \textbf{Cost/Hour (On-demand)} & \textbf{Cost Savings} \\
\hline
n2-standard-8 & \$0.078 & \$0.388 & 80\% \\
\hline
n2-standard-16 & \$0.156 & \$0.777 & 80\% \\
\hline
n2-standard-32 & \$0.311 & \$1.554 & 80\% \\
\hline
\end{tabular}
\end{table}

However, spot instances are not guaranteed to be available 100\% of the time. If your instance is interrupted, you will be given a few minutes to terminate your applications. To reduce the risk of your instance being interrupted, spot fleets can be used. A spot fleet is a group of spot instances that are launched together.

\subsubsection{Reserved Instances}\label{sec:cost-opt-reserved}
Reserved instances provide a discounted pricing model for enterprises that commit to using specific compute resources for a specified duration \cite{bib8} \cite{bib9} \cite{bib10}. By reserving instances in advance, organizations can secure a lower hourly rate compared to on-demand instances. Reserved instances are suitable for workloads with predictable and steady demand. Enterprises can choose from different reservation options, such as standard, convertible, or scheduled instances, based on their flexibility requirements.

\begin{table}
\centering
\caption{Cost savings with reserved instances vs. on-demand instances (us-east-1)}
\begin{tabular}{|c|c|c|c|}
\hline
\textbf{Instance Type} & \textbf{Cost/Hour (Reserved)} & \textbf{Cost/Hour (On-demand)} & \textbf{Cost Savings} \\
\hline
t2.large & \$0.0575 & \$0.0928 & 38.38\% \\
\hline
t2.medium & \$0.0287 & \$0.0464 & 39.53\% \\
\hline
t2.micro & \$0.0072 & \$0.0116 & 38.96\% \\
\hline
t2.nano & \$0.0036 & \$0.0058 & 39.66\% \\
\hline
t2.small & \$0.0144 & \$0.0230 & 40.00\% \\
\hline
t2.xlarge & \$0.1150 & \$0.1856 & 39.24\% \\
\hline
t3.large & \$0.0522 & \$0.0832 & 37.89\% \\
\hline
t3.medium & \$0.0261 & \$0.0416 & 38.75\% \\
\hline
t3.micro & \$0.0065 & \$0.0104 & 37.50\% \\
\hline
t3.small & \$0.0130 & \$0.0208 & 39.38\% \\
\hline
t3.xlarge & \$0.1043 & \$0.1664 & 37.67\% \\
\hline
\end{tabular}
\end{table}

\begin{table}
\centering
\caption{Cost savings with 1 year reserved instances vs. on-demand instances in Azure (eastus)}
\begin{tabular}{|c|c|c|c|}
\hline
\textbf{Instance Type} & \textbf{Cost/Hour (Reserved)} & \textbf{Cost/Hour (On-demand)} & \textbf{Cost Savings} \\
\hline
D2\_v5 & \$0.058 & \$0.096 & 40\% \\
\hline
D4\_v5 & \$0.115 & \$0.192 & 40\% \\
\hline
D8\_v5 & \$0.230 & \$0.384 & 40\% \\
\hline
D16\_v5 & \$0.461 & \$0.768 & 40\% \\
\hline
D32\_v5 & \$0.922 & \$1.536 & 40\% \\
\hline
\end{tabular}
\end{table}

\subsubsection{Serverless Computing}\label{sec:cost-opt-serverless}
Serverless computing eliminates the need for provisioning and managing servers. With serverless architectures, enterprises pay only for the actual compute time and resources used by their applications \cite{bib22}. This model offers granular cost control, as organizations are billed based on the number of function executions and resource consumption. By leveraging serverless computing, enterprises can optimize costs for event-driven workloads, where compute resources are only utilized when triggered by specific events. However, serverless computing may not be the cost efficient alternative to the equivalent compute offering if the application is long running \cite{bib23}.

\subsubsection{Containerization}\label{sec:cost-opt-container}
Containerization technologies, such as Docker and Kubernetes, provide efficient resource utilization by packaging applications and their dependencies into lightweight containers. Containerization enables enterprises to deploy applications consistently across different environments and scale them based on demand. By optimizing resource allocation and improving density, containerization can lead to cost savings by reducing the number of required compute instances. The table \ref{tab:container-cost} showcases the comparison of the cost incurred by traditional compute and containerized offerings for the specification of 2 vCPUs, 8 GB memory on a GCE instance type \texttt{e2-standard-2}, Cloud Functions function with similar specifications, and Cloud Run on GCP. All the pricing mentioned in the table \ref{tab:container-cost} is on-demand pricing in the region \texttt{us-central1}.

\begin{table}\label{tab:container-cost}
\centering
\caption{Comparison of containerized offerings vs. virtual machine on GCP}
\begin{tabular}{|c|c|c|}
\hline
\textbf{Compute offering} & \textbf{Type} & \textbf{Cost per hour} \\
\hline
GCE Instances & Virtual Machine & \$0.067006 \\
\hline
Cloud Functions & Serverless & \$0.0000068 \\
\hline
Cloud Run & Serverless (Containerized) & \$0.000018 \\
\hline
\end{tabular}
\end{table}

However, since Cloud Run is a serverless containerized offering from GCP, although it is the cheapest option to run containerized workloads, it can become quite expensive if the application is long-running and does not depend on event-driven architecture.

\subsubsection{VM Instance Types}\label{sec:cost-opt-instance-types}
One effective cost optimization strategy in the cloud is to leverage ARM-based instances instead of Intel or AMD-based instances. ARM-based processors, such as AWS Graviton (now in its fourth generation), Google Cloud's Tau T2A, and Azure's Ampere Altra-based instances, offer a compelling alternative in terms of cost efficiency. These processors are designed to deliver high performance while consuming less power, leading to lower operational costs \cite{bib24}. AWS Graviton3 (C7g, M7g, R7g instances) offers up to 25\% better compute performance than Graviton2, while Graviton4 (available in 2024) provides further improvements. By utilizing ARM-based instances, enterprises can typically achieve 20-40\% cost savings compared to equivalent x86 instances for compatible workloads. It's important to evaluate workload requirements and compatibility before migrating to ARM architecture, as some applications may require recompilation or have dependencies on x86-specific libraries. However, for compatible workloads including containerized applications, web servers, and many data processing pipelines migrating to ARM-based instances provides a cost-effective solution without compromising performance.

\begin{figure}[ht]
\centering
\begin{tikzpicture}[scale=1.2]
\begin{axis}[
    ybar,
    ymin=0,
    ymax=0.50,
    bar width=20pt,
    xlabel={Instance Type},
    ylabel={Cost per Hour (\$)},
    xtick=data,
    xticklabels={
        {Medium},
        {Large},
        {XLarge},
        {2XLarge}
    },
    nodes near coords,
    nodes near coords align={vertical},
    every node near coord/.append style={rotate=90, anchor=west, /pgf/number format/.cd, precision=4, fixed},
    enlarge x limits=0.2,
    tick style={draw=none},
    legend style={draw=none},
    ]
    \addplot[fill=red!40] coordinates {(1, 0.0416) (2, 0.0832) (3, 0.1664) (4, 0.3328)};
    \addplot[fill=blue!40] coordinates {(1, 0.0336) (2, 0.0672) (3, 0.1344) (4, 0.2688)};
    \legend{M6i Intel, M7g Graviton3}
\end{axis}
\end{tikzpicture}
\caption{Comparing the cost of Intel-based M6i instances vs. ARM-based M7g Graviton3 instances (Linux On-Demand Pricing, us-east-1)}
\label{fig:cost-comparison}
\end{figure}

\subsubsection{Idle Instances}\label{sec:cost-opt-idle}
Another effective strategy to reduce costs in cloud computing environments is through the implementation of automated discovery and deprecation of idle instances. Idle instances refer to virtual machines or cloud resources that are not actively utilized, yet still incur costs. By implementing a systematic approach to identify and list idle instances, organizations can gain visibility into their cloud usage patterns and identify opportunities for cost optimization.

The process may commence by monitoring resource usage and analyzing the utilization patterns of virtual machines. Through the use of monitoring tools and cloud management platforms, administrators can identify instances that consistently exhibit low or no utilization over a specified period. Once identified, these idle instances can be listed, allowing administrators to evaluate their necessity and potential for termination. By taking a proactive approach to manage idle instances, organizations can achieve significant cost savings. When an idle instance is listed, administrators have the opportunity to review its purpose and determine whether it is essential for ongoing operations. If the instance is found to be unnecessary or redundant, it can be safely stopped or terminated, eliminating the associated costs.

Implementing idle instance listing and stopping practices requires careful consideration of factors such as business requirements, service level agreements, and potential impacts on system performance. However, with proper planning and monitoring, organizations can achieve substantial cost savings while maintaining the required level of service availability.

\subsection{Storage}\label{sec:cost-opt-storage}
This section explores strategies and best practices for optimizing storage costs in a cloud environment. Efficient utilization of storage resources and implementing cost-saving techniques can help organizations reduce their storage expenses while ensuring data accessibility and reliability.

\subsubsection{Data Deduplication and Compression}\label{sec:cost-opt-data-dedup}
Data deduplication and compression techniques play a crucial role in optimizing costs in cloud storage environments. By eliminating redundant or unused data, organizations can significantly reduce storage requirements and associated expenses. Deduplication involves identifying and removing duplicate data segments, while compression reduces the size of data by encoding it using efficient algorithms. Together, these techniques offer substantial cost savings by minimizing storage capacity needs and mitigating the impact of data growth \cite{bib25}.

One of the key benefits of data deduplication is the elimination of redundant data copies. In many organizations, multiple users or applications may store identical or similar files, resulting in unnecessary data replication. By identifying and storing only unique data segments, deduplication reduces the overall storage footprint, leading to cost savings \cite{bib26}. Additionally, data compression techniques further enhance storage efficiency by reducing the size of individual files or data blocks. By employing compression algorithms, organizations can achieve significant data size reduction without compromising data integrity or accessibility.

The cost savings achieved through data deduplication and compression extend beyond storage capacity reduction. By minimizing the storage footprint, organizations can lower data transfer costs when moving data between cloud storage tiers or across different regions. The reduced data size also contributes to faster data transfer speeds, optimizing overall system performance. Moreover, data deduplication and compression techniques can enhance backup and disaster recovery processes, as smaller data volumes facilitate faster backup and recovery operations, reducing downtime and associated costs.

Implementing data deduplication and compression techniques requires careful consideration of factors such as data access patterns, application requirements, and computational overhead. It is crucial to select appropriate deduplication and compression algorithms that strike a balance between storage savings and processing overhead. Additionally, organizations must evaluate the impact on data access times and consider trade-offs between storage cost savings and the computational resources required for data deduplication and compression operations.


\subsubsection{Data Lifecycle Management Policies}\label{sec:cost-opt-dlm}

As data ages or becomes less frequently accessed, it may not require the same level of performance or accessibility as newer or more frequently used data. With data lifecycle management policies, organizations can define rules and criteria for data migration. This enables the automatic movement of data to lower-cost storage tiers, such as archival or cold storage, without sacrificing data availability or integrity \cite{bib27}.

By migrating data to more cost-effective storage options as it ages or becomes less frequently accessed, organizations can optimize their storage costs. This approach allows them to take advantage of different storage tiers that offer varying levels of performance, durability, and cost. It ensures that data is stored in the most suitable storage option while minimizing unnecessary expenses associated with storing all data in high-performance storage throughout its lifecycle.

Implementing data lifecycle management policies can be achieved through a combination of automated processes, data classification, and intelligent data management solutions. These policies can be tailored to specific business requirements, compliance regulations, and data access patterns \cite{bib28} \cite{bib29}. By adopting such policies, organizations can achieve significant cost savings by aligning storage costs with the value and usage patterns of their data.

\begin{table}
\centering
\caption{Comparison of Data Storage Tiers in Google Cloud Storage (Cost per GB per Month)}
\begin{tabular}{|c|c|c|c|c|}
\hline
Location & Standard & Nearline & Coldline & Archive \\
\hline
Iowa (us-central1) & \$0.020 & \$0.010 & \$0.004 & \$0.0012 \\
\hline
Frankfurt (europe-west3) & \$0.023 & \$0.013 & \$0.006 & \$0.0025 \\
\hline
Tokyo (asia-northeast1) & \$0.023 & \$0.016 & \$0.006 & \$0.0025 \\
\hline
Sydney (australia-southeast1) & \$0.023 & \$0.016 & \$0.006 & \$0.0025 \\
\hline
São Paulo (southamerica-east1) & \$0.035 & \$0.020 & \$0.007 & \$0.0030 \\
\hline
\end{tabular}
\end{table}

\begin{figure}[ht]
    \centering
    \begin{tikzpicture}
        \begin{axis}[
            xlabel={Location},
            ylabel={Cost per GB per Month (\$)},
            symbolic x coords={Iowa, Frankfurt, Tokyo, Sydney, São Paulo},
            xtick=data,
            ymin=0,
            ymax=0.04,
            legend style={at={(0.5,-0.2)},anchor=north},
            width=\textwidth,
            height=8cm,
            grid=major,
        ]
        
        \addplot[blue, mark=*] coordinates {
            (Iowa, 0.020)
            (Frankfurt, 0.023)
            (Tokyo, 0.023)
            (Sydney, 0.023)
            (São Paulo, 0.035)
        };
        \addplot[red, mark=triangle*] coordinates {
            (Iowa, 0.010)
            (Frankfurt, 0.013)
            (Tokyo, 0.016)
            (Sydney, 0.016)
            (São Paulo, 0.020)
        };
        \addplot[green, mark=diamond*] coordinates {
            (Iowa, 0.004)
            (Frankfurt, 0.006)
            (Tokyo, 0.006)
            (Sydney, 0.006)
            (São Paulo, 0.007)
        };
        \addplot[olive, mark=square*] coordinates {
            (Iowa, 0.0012)
            (Frankfurt, 0.0025)
            (Tokyo, 0.0025)
            (Sydney, 0.0025)
            (São Paulo, 0.0030)
        };
        
        \legend{Standard, Nearline, Coldline, Archive}
        \end{axis}
    \end{tikzpicture}
    \caption{Comparison of Data Storage Tiers in Google Cloud Storage}
    \label{fig:storage-tiers}
\end{figure}

\subsubsection{Data Archiving and Retention}\label{sec:cost-opt-archive}
Another important tactic for cloud storage cost optimization is enacting a data archival and retention policies. Enterprises are often required to retain data for compliance which can result in expensive cloud storage costs, especially if that data is kept in high-performance storage tiers. Implementing data archiving and retention standards becomes crucial to overcoming this problem.

To achieve cost efficiency, organizations can leverage specific storage options tailored for long-term data retention, such as archive storage tiers. These tiers offer significantly lower storage costs compared to standard storage tiers, while still ensuring sufficient durability and availability. Based on the data governance policies, the data can be migrated from high-performance tier to archive tier. Additionally, setting a small retention value on the data would result in the smaller data storage footprint for an enterprise resulting in significant cost savings in data storage.

By harnessing intelligent data management techniques, organizations can identify and apply suitable retention periods to different data sets, ensuring compliance with legal requirements while optimizing storage costs.

\subsection{Network}\label{sec:cost-opt-network}
Optimizing the networking cost is essential since data transport and communication between different components can account for a sizable portion of overall cloud costs. This section looks at many strategies and best practices that businesses can use to reduce network expenses while maintaining top performance. This section offers suggestions for enhancing the efficiency of networking infrastructure within the cloud, from pattern analysis and optimization to traffic control and efficient network service utilization.

\subsubsection{Optimizing Network Traffic Patterns}\label{sec:cost-opt-traffic}
In order to find chances for cost reductions, it is essential to examine the network traffic patterns inside a cloud environment. Enterprises may optimize their network infrastructure and cut expenses by looking at the amount and kind of data flows using network traffic analysis. Enterprises can use monitoring tools and analytics platforms offered by cloud providers or third-party solutions to do network traffic analysis. These tools give users insight into the patterns of network traffic, including data transfer rates, peak usage periods, and the types of data being transferred \cite{bib30}.

Enterprises can find potential areas for optimization with the use of this information. For instance, they can discover data-intensive programs or services that cause a lot of network traffic and look for ways to improve how data is transferred between them. This may entail putting data compression techniques into practice, leveraging content delivery networks (CDNs) or edge caching to shorten the distances between data transfers, or using data deduplication techniques to stop repeated transfers.

In order to maximize network efficiency, network traffic analysis can also help find opportunities for traffic rerouting or load balancing. Enterprises can lower bandwidth consumption and possibly lower data transfer costs by intelligently directing traffic through efficient channels or dispersing it across numerous network resources.

\subsubsection{Content Delivery Networks (CDNs) and Edge Caching}\label{sec:cost-opt-cdn}
Content Delivery Networks (CDNs) play a key role in improving network performance and lowering costs. When using CDNs, content can be delivered from the edge location that is closest to the end users by utilizing edge servers that are dispersed across different locations. The usage of CDNs, edge caching, and traffic control strategies to save costs and boost overall network effectiveness are examined in this section. By caching and distributing content closer to end users, CDNs aim to lower latency and boost speed. CDNs reduce the distance that data must travel across the network by strategically distributing content in geographically dispersed edge servers. By serving content from edge locations rather than the origin server, this not only improves the user experience but also lowers network egress costs.

Frequently accessed content is stored at edge server locations as part of the edge caching approach. The content is delivered directly from the nearest edge cache when a user wants it, avoiding the need for data to travel across the entire network. Enterprises may drastically lower network egress costs and enhance end-user response times by utilizing edge caching. Additionally, by offloading the traffic from the origin server to the edge locations, enterprises can lessen the stress on the origin server, increase network bandwidth, and lower egress costs by dumping data closer to the end users. As a result, less expensive origin server resources and data transfer are not required for serving static content.

Finally, optimizing network utilization and cutting costs need effective traffic engineering. To achieve the best distribution of network traffic, a network engineer may use smart load balancing and intelligent routing algorithms. Enterprises can reduce network egress and ingress expenses by routing traffic through the most economical pathways in each of cloud regions. In order to increase performance and save costs, traffic engineering techniques can also be used to prioritize important traffic, reduce bottlenecks, and optimize bandwidth use \cite{bib31}.

\subsubsection{Minimizing Data Transfer Size}\label{sec:cost-opt-data-transfer}
Utilizing data compression techniques to limit pointless data transfer is one efficient strategy. Data compression can drastically reduce the quantity of data carried over the network, resulting in lower network bandwidth usage and cheaper expenses. The type of data being compressed and the chosen compression ratio will determine the optimal approach to compress the data before transmitting it over the network. For example, the best compression ratio will typically come from lossless compression, but it will also cause some delay. On the other hand, lossy compression will result in some quality loss but can offer a better compression ratio with less latency. 

Twitter's (now X) Parquet encoding standard is one such example of using compression techniques for data that is being backed up to the cloud. The capability of Parquet compression to compress data at the column level is one of its main benefits \cite{bib32}. By utilizing similarities and redundancies among columns, this columnar storage strategy enables effective compression, producing higher compression ratios than row-based storage formats. Twitter was able to improve storage expenses in their cloud environment by lowering the size of the data saved in Parquet files. The adoption of Parquet compression by Twitter emphasizes the significance of choosing compression methods that are suited to the data and use cases. 


\subsubsection{Network Tuning}\label{sec:cost-opt-tuning}
In large cloud environments, optimizing network configurations is a crucial part of reducing costs. Enterprises can maximize network performance, decrease data transfer, and cut associated expenses by carefully tuning the network settings for their compute infrastructure. One common approach is using load balancing and efficient routing algorithms to evenly divide network traffic across the available resources and prevent pointless data transfer. In addition, streamlining network protocols and configurations, like TCP/IP settings, can increase network effectiveness and cut down on bandwidth usage. Prioritizing vital network traffic and efficiently allocating bandwidth resources can be achieved by using traffic shaping and quality of service (QoS) regulations. Enterprises can also use network analytics and monitoring technologies to get insights into network usage trends and spot optimization opportunities. Therefore, it is possible to significantly reduce network expenses while retaining optimal performance and reliability by routinely analyzing and fine-tuning network configurations in accordance with shifting workload demands and cost optimization objectives.

\subsection{Logging}\label{sec:cost-opt-logging}
Cloud Logging plays a crucial role in managing and monitoring the vast amounts of logs generated by cloud-based applications and infrastructure. While logging is necessary for troubleshooting, performance analysis, and compliance, if it is not managed effectively, it can also result in considerable expenses. This section examines various cost-cutting and cost-optimization techniques for cloud logging.

\subsubsection{Log Filtering}\label{sec:cost-opt-filtering}
Implementing efficient log filtering and sampling methods is the first step in managing logging costs effectively. Enterprises can reduce the amount of logs stored and transmitted by defining precise filters, concentrating only on pertinent data. For example, log severity, particular components or services, or user-defined criteria can all be used as filters. Similar to this, log sampling enables the collection of a representative subset of logs as opposed to storing each individual log entry. Enterprises can strike a balance between cost reduction and keeping a sufficient level of system visibility by carefully choosing the sampling rate. For example, let's say that a company stores 100 GB of log data per month. If the company does not use log filtering, then all of this log data will be stored. However, if the company uses log filtering to only store logins, errors, and other important events, then the amount of log data that is stored can be reduced to 10 GB per month. This would save the company 90\% on storage costs.

\subsubsection{Log Storage}\label{sec:cost-opt-log-storage}
Utilizing data compression methods and picking the best log storage option can have a big impact on cost reduction. Different storage tiers, such as standard storage, cold storage, or archival storage, are available from cloud providers at various price points. Enterprises can choose which logs to store in the most economical storage tier by analyzing the frequency and urgency of log access. Additionally, as mentioned in the section of data compression, implementing compression techniques for the logging data can reduce log size and minimize the storage costs without compromising the logs. For example, Twitter's (now X) use of LZO compression to compress Scribe event log data \cite{bib33} and Meta's use of ZStandard library to compress live logging data \cite{bib34} are two prominent examples of how using compression techniques in log storage can optimize the cost for a company's infrastructure \cite{bib35}.

\subsubsection{Log Retention}\label{sec:cost-opt-retention}
Another major expense incurred by enterprises is retaining logs for longer periods of time. When storing logs for an extended period, especially less important or logs related to regulatory compliance, extra costs may arise. By leveraging a data lifecycle management policy, the process of archiving or deleting logs in accordance with predefined policies can be automated thereby reducing the human intervention. Therefore, logs can be retained for the exact amount of time that is required by setting retention periods based on compliance requirements, business needs, and industry best practices. Moreover, by combining the log retention policy with the techniques like data compression and choosing an appropriate storage tiers for compliance-related logs, enterprises can incur a substantial amount of cost savings. Additionally, based on the contracts between cloud provider and enterprises, enterprises should elect for the cheapest storage solution to store the logging data.


\subsubsection{Log Monitoring and Alerting}\label{sec:cost-opt-alering}
To effectively optimize costs, log usage and costs must be proactively tracked. To track log volume, storage usage, and associated costs, organizations should set up alerts and notifications. Organizations can spot any unexpected spikes or patterns in log volume and take prompt action to minimize costs by setting thresholds and proactive monitoring mechanisms. Cost effectiveness is maintained over time by periodically reviewing log usage patterns and modifying monitoring strategies in response to changing needs.

\subsection{Resource Recommendations}\label{sec:cost-opt-reco}
To help businesses reduce their cloud costs, cloud providers provide a variety of tips and tools in form of Resource Recommendations. These suggestions are supported by resource configurations, usage trends, and industry best practices. Enterprises can identify potential cost-saving opportunities, maximize resource usage, and align their cloud spending with business goals by utilizing these insights. The significance of cloud provider recommendations is discussed in this section, along with the important areas where they can be used to save money.

\subsubsection{Compute Instance Recommendations}\label{sec:cost-opt-reco-compute}
Cloud providers offer recommendations for utilizing reserved instances (RIs) and savings plans to optimize costs. These recommendations are usually focused on compute instances like Amazon AWS EC2. For example, AWS provides recommendations for utilizing Reserved Instances (RIs) and Savings Plans based on usage patterns and potential cost savings. Whereas, GCP provides recommendations for using the Commited Use Discounts (CUDs) to optimize costs for the long-term contracts and Azure offers Reserved VM Instances (RIs) and reservation recommendations for the same purposes, respectively. In addition to the long-term commitments, these compute instance recommendations also involve offering the list of idle instances, underutilized, or over-provisioned resources. For example, GCP offers a machine learning-based Recommenders that observe the enterprises' virtual machine instances for 8 days and then offer the recommendations to right-size the instances to save costs \cite{bib36}.


\subsubsection{Cloud Storage Recommendations}\label{sec:cost-opt-reco-storage}
Cloud service providers provide recommendations for maximizing storage usage and costs. These suggestions look at storage usage patterns, point out ineffective or unused storage resources, and make suggestions for the best storage configurations. Enterprises can reduce wasteful storage expenses, improve data placement, and take advantage of cost-efficient storage tiers by implementing these recommendations. All major cloud providers offer cloud stoarge recommendations to help save storage costs. For example, AWS offers S3 Storage Lens to analyze and optimize storage usage, along with Amazon S3 Intelligent-Tiering for automated data tiering recommendations \cite{bib28}. On the other hand, GCP provides a feature called GCS Autoclass. Based on each object's access pattern, the Autoclass feature automatically moves objects in the bucket to the proper storage classes \cite{bib27}. This feature moves data that is frequently accessed to Standard storage to improve future accesses and moves data that is not accessed to colder storage classes facilitating automated data lifecycle management to save costs.


\subsubsection{Database Recommendations}\label{sec:cost-opt-reco-db}
Cloud service providers also offer recommendations for improving database configurations and usage. These suggestions examine database usage patterns, query efficiency, and performance, and finally they offer optimization tactics. Organizations can optimize database costs, reduce wasteful resource use, and improve database performance by putting these recommendations into practice. Such examples of database recommendations can be found in AWS' Amazon RDS Performance Insights and Database Query Monitoring for identifying and optimizing database performance issues, Azure's SQL Database Advisor \cite{bib38}, and finally GCP's  CloudSQL Insights that also offers recommendations on idle disks as well as over-provisioned CloudSQL instances \cite{bib37}.


\subsubsection{Cloud Network Recommendations}\label{sec:cost-opt-reco-nw}
The cloud network recommendations provided by the cloud providers mainly focus on the idle resources that still have IP addresses attached to them. For example, GCP's Idle Resource Recommender would identify resources like persistent disks (PDs), IP addresses, and custom disk images that aren't used. Since the IP addresses are pay-per-use resources in the cloud, deleting an instance or releasing the IP address from that instance would save 100\% of the cost associated with the IP addresses. On the other hand, Amazon VPC IP Address Manager (IPAM) offered by AWS aids in managing an organization's IP inventory. Therefore, with the help from IPAM, an enterprise can identity the idle IP addresses and release those IP addresses from the compute resources to save costs \cite{bib39}.


\subsection{Committed Use Discounts}\label{sec:cost-opt-cud}
Cloud service providers like Microsoft Azure, Amazon Web Services, and Google Cloud Platform (GCP) offer committed use discounts (CUDs) as a way to cut costs. Customers who use CUDs agree to use a certain number of cloud resources (such as compute instances, storage, or databases) for a predetermined period of time, usually one or three years. In turn, cloud service providers offer significant discounts on the hourly rates of the committed resources in exchange for this commitment. Enterprises can significantly reduce their cloud infrastructure costs by utilizing CUDs, especially for long-term workloads with predictable usage patterns. While still enjoying the scalability and flexibility of cloud computing, this pricing model enables organizations to effectively plan and budget their cloud costs. Additionally, some cloud providers also offer flexibility in terms of instance family, region, and instance size within the committed use, providing customers with options to optimize their usage further. CUDs are an effective cost optimization strategy for enterprises seeking long-term cloud resource utilization and cost predictability.

Let's consider the on-demand price for GCP's n2-standard-16 machine type with 16 vCPUs and 64GB memory in the us-central1 region, which is \$0.776944 per vCPU hour. Assuming daily usage of 1000 vCPU hours, we will determine the cost savings with Google Cloud Platform's (GCP) Committed Use Discounts (CUDs) for a 1-year and 3-year commitment. For a 1-year commitment with a 28\% discount, the hourly rate would be reduced to \$0.55964832. With a daily usage of 1000 vCPU hours, the daily cost would be \$559.65. Over the course of a year, this would result in a total cost of \$204,012.1. Compared to the on-demand cost of \$283,558.4, this represents a savings of \$79,546.31. On the other hand, For a 3-year commitment with a 46\% discount, the hourly rate would be further reduced to \$0.41999824. With a daily usage of 1000 vCPU hours, the daily cost would be \$420.1. Over three years, this would result in a total cost of \$459,898.97. Compared to the on-demand cost, this represents a savings of \$177,659.44.

These calculations show the potential cost savings that can be realized when using the n2-standard-16 machine type in the us-central1 region with GCP's Committed Use Discounts. It's important to note that the precise commitment terms, usage trends, and instance types selected will determine the actual savings. Nevertheless, using CUDs can greatly lower the overall cost of running compute workloads in the cloud, making it an affordable choice for businesses using the GCP infrastructure. Similarly, AWS and Microsoft Azure offer similar CUDs for the long-term commitments on various cloud resources.

\subsection{System Rearchitecture}\label{sec:cost-opt-rearch}
The infrastructure or system rearchitecture is a strategic approach to improve cloud infrastructure's cost effectiveness. Organizations can find opportunities to cut costs while improving performance, scalability, and reliability by reevaluating the system's design and structure. This subsection explores some key areas where re-architecture can lead to significant cost savings.

\subsubsection{Microservices vs. Monolithic Architecture}\label{sec:cost-opt-micro-monolith}
Organizations can break down monolithic applications into more manageable, independent services by implementing a modularized and microservices architecture. This architectural strategy has several cost-cutting advantages. First, it enables granular scaling, which prevents overprovisioning of resources by only scaling the required services in accordance with demand. Additionally, by precisely allocating resources to each service, microservices encourage efficient resource utilization while lowering overall infrastructure costs. By decoupling services, organizations can also take advantage of different pricing models, such as serverless computing, paying only for actual usage and achieving cost optimization.

Monolithic architecture, on the other hand, may provide management simplicity and potential cost savings as showcased in \cite{bib40}. Organizations with monolithic architectures have a single code base, which lessens the challenges of managing and coordinating numerous microservices. Less resources are needed for monitoring, testing, and maintaining a single application, which can result in lower development, deployment, and operational costs. Additionally, because the entire application can run on a limited number of servers or containers, monolithic architecture may require less infrastructure resources than a microservices architecture, which lowers hosting and scaling costs. However, it is important to assess the specific needs and goals of the organization, as well as the scalability and future growth considerations, before deciding on the most suitable architectural approach for cost optimization.

\subsubsection{Replacing Virtual Machines with Containers}\label{sec:cost-opt-vm-container}
Containerization technologies like Docker, coupled with orchestration frameworks such as Kubernetes, offer cost optimization benefits by improving resource utilization and workload management. Containers provide lightweight and isolated environments for applications, reducing the overhead of running multiple virtual machines. Organizations can effectively manage the deployment, scaling, and monitoring of containers with container orchestration, maximizing resource utilization and cutting costs. Additionally, because multiple containers can be installed on a single virtual machine, containerization enables more effective use of cloud resources while reducing infrastructure and licensing costs.

\subsubsection{Autoscaling Infrastructure}\label{sec:cost-opt-autoscaling-infra}
Cost optimization calls for the capacity to scale resources automatically in response to demand. Organizations can dynamically modify their infrastructure to suit workload patterns by implementing autoscaling policies. By ensuring that resources are only provisioned when necessary, autoscaling helps to cut costs during times of low demand. Cloud providers offer various autoscaling mechanisms, such as scaling based on CPU utilization, network traffic, or custom metrics, allowing organizations to right-size their infrastructure and optimize costs.

\subsubsection{Serverless Computing}\label{sec:cost-opt-faas}
In serverless computing, also referred to as Function-as-a-Service (FaaS), programmers concentrate on writing code for particular functions rather than managing or setting up servers. The expense and difficulty of managing unused or underutilized resources are eliminated by this paradigm shift. With serverless, businesses save a lot of money by only paying for the time that functions actually take to execute. Therefore, by leveraging auto-scaling capabilities provided by the cloud provider, organizations can effortlessly handle workload fluctuations without incurring additional costs associated with idle resources.

These examples show how re-architecting a system or infrastructure can lead to significant cost savings. Re-architecting a system, however, necessitates careful planning, in-depth research, and an in-depth understanding of the current infrastructure and business requirements. Organizations can achieve cost optimization while enhancing agility, scalability, and resilience in the cloud by utilizing the advantages of modularization, microservices, serverless computing, containerization, and autoscaling.

\section{AI and ML Infrastructure Cost Optimization}\label{sec:ai-cost-opt}
The rapid adoption of artificial intelligence and machine learning has introduced new cost optimization challenges and opportunities. Unlike traditional cloud workloads, AI/ML workloads are characterized by intensive GPU utilization, large model sizes, and distinct phases (training vs. inference) with different resource requirements. This section explores emerging techniques for optimizing AI infrastructure costs.

\subsection{The Economics of AI Infrastructure}\label{sec:ai-economics}
The AI infrastructure market has experienced explosive growth, reaching \$50 billion in 2024 with projected annual growth of 35\% through 2027 \cite{bib45}. For organizations building AI applications, GPU compute typically represents 40-60\% of technical budgets in the first two years of operation. Understanding the cost structure of AI workloads is essential for effective optimization.

AI infrastructure costs can be categorized into several key components. Compute resources, particularly GPUs and TPUs, represent the primary cost driver, with high-end GPUs like NVIDIA H100 commanding \$2-5 per GPU-hour and the newer H200 reaching \$3.50-6.50 per GPU-hour depending on the provider and commitment level \cite{bib47}. Storage requirements for model checkpoints, training datasets, and inference logs can be substantial, particularly for large-scale training runs that may generate terabytes of checkpoint data. Networking costs arise from data transfer between GPUs in distributed training, cross-region replication, and serving inference traffic globally, with egress fees often surprising organizations that underestimate data movement patterns. Finally, model serving requires persistent compute resources that must be scaled to match traffic patterns, presenting challenges similar to traditional web application scaling but with the added complexity of GPU resource management.

\subsection{GPU Instance Selection and Pricing}\label{sec:gpu-pricing}
Selecting the appropriate GPU instance type is critical for AI cost optimization. Cloud providers offer various GPU options with different price-performance characteristics.

\subsubsection{GPU Pricing Landscape}
As of 2025, the major cloud providers offer several GPU instance types for AI workloads. AWS provides P4d instances (NVIDIA A100), P5 instances (NVIDIA H100), and P5en instances (NVIDIA H200). In June 2025, AWS reduced prices for GPU instances by up to 45\%, with H100 instances seeing a 44\% reduction in on-demand pricing \cite{bib48}. Google Cloud offers A2 instances (A100) and A3 instances (H100), while Microsoft Azure provides ND-series instances with similar GPU configurations.

\begin{table}[ht]
\centering
\caption{Approximate GPU Cloud Pricing Comparison (2025)}
\label{tab:gpu-pricing}
\small 
\begin{tabularx}{\textwidth}{|X|c|c|c|} 
\hline
\textbf{GPU Type} & \textbf{AWS (\$/hr)} & \textbf{GCP (\$/hr)} & \textbf{Azure (\$/hr)} \\
\hline
NVIDIA A100 (80GB) & \$2.50--3.50 & \$2.80--3.60 & \$2.70--3.50 \\
\hline
NVIDIA H100 & \$3.50--5.00 & \$4.00--5.50 & \$4.00--5.50 \\
\hline
NVIDIA H200 & \$4.50--6.00 & \$5.00--6.50 & \$5.00--6.50 \\
\hline
\end{tabularx}
\end{table}

Specialized GPU cloud providers (often called ``neoclouds'') such as Lambda Labs, CoreWeave, and others offer competitive pricing, sometimes 30-50\% lower than hyperscalers for equivalent hardware \cite{bib47}. However, these providers may offer fewer integrated services and less geographic coverage.

\subsubsection{GPU Reserved Instances and Savings Plans}
Similar to traditional compute instances, GPU instances can be reserved for significant discounts. AWS Savings Plans for GPU instances offer 25-45\% discounts for 1-3 year commitments \cite{bib48}. Organizations with predictable AI training schedules should evaluate reserved capacity to reduce costs.

\subsection{LLM Inference Cost Optimization}\label{sec:llm-inference}
Large language model inference has emerged as a significant cost center for organizations deploying AI applications. However, the economics of LLM inference have improved dramatically since 2021.

\subsubsection{The LLMflation Phenomenon}
According to analysis by Andreessen Horowitz, LLM inference costs have decreased by approximately 10x annually since the public introduction of GPT-3 in 2021 \cite{bib46}. What cost \$60 per million tokens in November 2021 now costs approximately \$0.06 per million tokens for equivalent model performance. This 1000x cost reduction over four years has been driven by several converging factors. Hardware improvements through new GPU architectures such as the H100 and H200 offer substantially better cost-performance ratios than their predecessors. Model quantization techniques that reduce precision from 16-bit to 8-bit or 4-bit have significantly decreased compute and memory requirements without proportional quality loss. The development of smaller, more efficient models means that modern 1-billion parameter models can exceed the performance of 175-billion parameter models from 2021 on many benchmarks. Software optimizations including Flash Attention, speculative decoding, and continuous batching have reduced computational overhead substantially. Finally, open source competition from models released by Meta (Llama series), Mistral, and Chinese providers like DeepSeek has intensified price competition and compressed margins across the industry.

\begin{figure}[ht]
\centering
\begin{tikzpicture}
\begin{axis}[
    xlabel={Date},
    ylabel={Cost per Million Tokens (\$)},
    ymode=log,
    xmin=0,
    xmax=5,
    ymin=0.01,
    ymax=10000,
    xtick={0,1,2,3,4,5},
    xticklabels={Nov 2021, Nov 2022, Nov 2023, Nov 2024, Nov 2025, Aug 2025},
    legend pos=north east,
    grid=major,
]

\addplot[blue, mark=*, thick] coordinates {
    (0, 60)
    (1, 6)
    (2, 0.6)
    (3, 0.15)
    (4, 0.06)
    (5, 0.05)
};
\addlegendentry{MMLU 42 (GPT-3 level)}

\addplot[red, mark=triangle*, thick, dashed] coordinates {
    (2, 60)
    (3, 10)
    (4, 1)
    (5, 0.25)
};
\addlegendentry{MMLU 83 (GPT-4 level)}

\addplot[green!60!black, mark=square*, thick, dotted] coordinates {
    (4, 10)
    (5, 1.25)
};
\addlegendentry{MMLU 88+ (GPT-5 level)}

\end{axis}
\end{tikzpicture}
\caption{LLM Inference Cost Decline (2021-2025) for Equivalent Model Performance}
\label{fig:llm-cost-decline}
\end{figure}

\subsubsection{LLM API Pricing Evolution}
The pricing history of major LLM APIs illustrates the rapid cost decline in frontier models \cite{bib49}. When OpenAI launched GPT-4 in March 2023, pricing stood at \$30/\$60 per million tokens for input and output respectively. By November 2023, GPT-4 Turbo reduced this to \$10/\$30 per million tokens. The introduction of GPT-4o in May 2024 brought prices down to \$5/\$15 per million tokens, and GPT-4o Mini in July 2024 offered dramatically lower rates at \$0.15/\$0.60 per million tokens for simpler use cases.

The trend has continued into 2025, with OpenAI releasing GPT-5 in August 2025 at \$1.25/\$10 per million tokens for input and output, representing a significant capability improvement at lower cost than the original GPT-4. The GPT-5 Mini variant offers even more aggressive pricing at \$0.25/\$2 per million tokens, while GPT-5 Nano targets high-volume applications at \$0.05/\$0.40 per million tokens. Anthropic's Claude models follow similar tiered pricing, with Claude Opus 4.1 at \$15/\$75 per million tokens for the most capable tier, Claude Sonnet 4 at \$3/\$15 per million tokens for balanced performance, and Claude Haiku 3.5 at \$0.80/\$4 per million tokens for cost-sensitive applications. Chinese providers like DeepSeek have further intensified price competition, offering capable models at a fraction of Western provider costs and triggering what analysts describe as a shift from a performance race to a price war.

This pricing evolution represents approximately a 95\% reduction in costs for equivalent model capabilities over the three-year period from 2023 to 2026, fundamentally changing the economics of AI-powered applications.

\subsection{Model Quantization}\label{sec:quantization}
Model quantization is a technique that reduces the precision of model weights and activations from higher bit formats (e.g., 32-bit or 16-bit floating point) to lower bit formats (e.g., 8-bit or 4-bit integers). This reduction significantly decreases memory requirements and computational costs while maintaining acceptable model quality.

\subsubsection{Quantization Schemes}
Research from Red Hat and Neural Magic demonstrates that quantized models can achieve near-full accuracy recovery across various benchmarks \cite{bib50}. The W8A8-INT scheme quantizes both weights and activations to 8-bit integers, providing approximately 2x model size compression and 1.8x inference speedup, making it ideal for server deployments on NVIDIA Ampere (A100) and older hardware. The W8A8-FP scheme uses 8-bit floating point format for weights and activations, offering similar compression and speedup characteristics while being optimized for NVIDIA Hopper (H100) and Ada Lovelace hardware. For more aggressive optimization, the W4A16-INT scheme quantizes weights to 4-bit integers while maintaining activations at 16-bit precision, achieving approximately 3.5x model size compression and 2.4x speedup for latency-critical applications.

\begin{table}[ht]
\centering
\caption{Quantization Impact on Model Size and Performance}
\begin{tabular}{|c|c|c|c|}
\hline
\textbf{Scheme} & \textbf{Size Reduction} & \textbf{Speedup} & \textbf{Best Use Case} \\
\hline
W8A8-INT & 2x & 1.8x & Server/throughput workloads \\
\hline
W8A8-FP & 2x & 1.8x & H100/Ada hardware \\
\hline
W4A16-INT & 3.5x & 2.4x & Edge/latency-critical \\
\hline
\end{tabular}
\label{tab:quantization}
\end{table}

\subsection{Inference Optimization Techniques}\label{sec:inference-opt}
Beyond quantization, several techniques can substantially reduce LLM inference costs for production deployments.

\subsubsection{Batch Processing}
OpenAI and other providers offer batch APIs that process requests asynchronously at 50\% lower cost \cite{bib49}. The OpenAI Batch API provides a 24-hour turnaround time with significantly higher rate limits, making it ideal for non-time-sensitive workloads such as data processing, evaluations, and content generation. Anthropic offers similar batch pricing with 50\% discounts on both input and output tokens for asynchronous processing.

\subsubsection{Model Selection and Routing}
Not all queries require frontier model capabilities. Implementing intelligent routing that directs simple queries to smaller, cheaper models can reduce costs by 90\% or more for appropriate workloads. For example, routing straightforward queries to GPT-5 Nano at \$0.05/million input tokens instead of GPT-5 at \$1.25/million represents a 25x cost reduction. Organizations increasingly adopt tiered model strategies, reserving expensive frontier models like Claude Opus 4.1 or GPT-5 for complex reasoning tasks while handling routine queries with efficient models like Claude Haiku or GPT-5 Nano.

\subsubsection{Caching and Semantic Deduplication}
Caching responses for repeated or semantically similar queries can eliminate redundant inference costs. Techniques include exact match caching for identical queries, semantic similarity caching that identifies functionally equivalent requests, and prompt prefix caching supported natively by providers like Anthropic, where cached input tokens cost only 10\% of standard rates. For applications with repetitive query patterns, effective caching strategies can reduce inference costs by 50-80\%.

\subsubsection{Context Window Optimization}
LLM costs scale directly with context length, making context optimization essential for cost control. Summarizing long documents before inclusion in prompts can reduce token counts by 80-90\% while preserving essential information. Retrieval-augmented generation (RAG) architectures include only relevant context chunks rather than entire documents, dramatically reducing per-request token usage. For conversational applications, implementing sliding window approaches for conversation history prevents context from growing unboundedly while maintaining coherent interactions.

\subsubsection{Cost-Aware LLM Applications}
Beyond optimizing the LLM inference itself, organizations must consider the downstream costs of LLM-generated outputs. A study on cost-aware Text-to-SQL systems demonstrates that LLM-generated queries can exhibit significant cost variance when executed on cloud data warehouses \cite{bib56}. Evaluating six state-of-the-art LLMs across 180 query executions on Google BigQuery, the research found that reasoning models such as o1-preview process 44.5\% fewer bytes than standard models while maintaining equivalent correctness rates of 96.7\%-100\%. Notably, execution time correlates weakly with query cost (r=0.16), indicating that optimizing for speed does not necessarily optimize for cost. Models exhibit up to 3.4x cost variance, with standard models producing outliers exceeding 36GB per query due to inefficiency patterns such as missing partition filters and unnecessary full-table scans. This research highlights the importance of evaluating not just LLM inference costs, but also the cloud compute costs incurred by LLM-generated outputs in enterprise environments. Organizations deploying Text-to-SQL or similar LLM applications should implement cost monitoring for downstream query execution and consider using reasoning models for cost-sensitive workloads.

\subsection{Training Cost Optimization}\label{sec:training-opt}
While inference costs dominate for deployed applications, training costs remain significant for organizations developing custom models.

\subsubsection{Spot and Preemptible Instances}
Training workloads that can checkpoint and resume are well-suited for spot instances, which offer 60-90\% discounts. Frameworks like PyTorch and TensorFlow support checkpointing, enabling cost-effective training on interruptible capacity.

\subsubsection{Mixed Precision Training}
Using mixed precision (FP16 or BF16) instead of FP32 can reduce training time and memory usage by approximately 2x with minimal impact on model quality. Modern frameworks and hardware (NVIDIA Tensor Cores) are optimized for mixed precision operations.

\subsubsection{Efficient Fine-tuning}
Parameter-efficient fine-tuning techniques such as LoRA (Low-Rank Adaptation) and QLoRA reduce the computational requirements for adapting pre-trained models. These techniques can reduce fine-tuning costs by 10-100x compared to full model fine-tuning while achieving comparable results.

\subsection{FinOps for AI}\label{sec:finops-ai}
The FinOps Foundation has recognized AI cost management as a distinct discipline, introducing ``FinOps for AI'' as a focus area \cite{bib51}. This emerging practice encompasses several key capabilities that organizations must develop to effectively manage AI infrastructure costs. GPU utilization monitoring enables teams to track utilization rates and identify idle or underutilized resources, which is particularly important given the high hourly costs of GPU instances. Cost allocation by model and experiment allows organizations to attribute costs to specific models, experiments, or teams, providing the visibility needed for informed decision-making about AI investments. Inference cost tracking monitors per-request costs and identifies optimization opportunities, enabling teams to understand the true cost of serving AI-powered features. Capacity planning forecasts GPU requirements based on model deployment schedules and anticipated demand, helping organizations balance cost efficiency with performance requirements.

Tools like Kubecost and OpenCost (promoted to CNCF Incubation status in October 2024) have expanded their capabilities to support GPU cost allocation in Kubernetes environments \cite{bib52}. These tools provide visibility into GPU utilization, cost attribution by namespace and workload, and integration with cloud provider billing data, enabling organizations to apply the same FinOps rigor to AI workloads that they apply to traditional cloud infrastructure.

\section{Case Studies}\label{sec:case-studies}
The application of cost optimization techniques in real-world scenarios is crucial for businesses seeking to maximize their efficiency and minimize infrastructure expenses. In this section, we examine case studies of well-known businesses that have successfully implemented cost optimization strategies, including Amazon Prime Video, Pinterest, Baselime (acquired by Cloudflare), and Netflix. These case studies span from 2023 to 2025 and offer insightful information about the practical application of the various techniques covered in this paper, demonstrating how architectural decisions, resource optimization, platform migration, database consolidation, and strategic technology choices can lead to significant cost savings ranging from 28\% to over 90\%. We can learn important lessons and best practices for efficient cost optimization by looking at how these organizations handled their infrastructure challenges and realized significant cost savings. We aim to demonstrate the variety of strategies used by businesses to cut costs while maintaining or even improving the performance and scalability of their infrastructure through these real-world examples.

\subsection{Amazon Prime Video}\label{sec:prime}
Prime Video is a video streaming service that offers a wide variety of movies and TV shows. This section discusses how Prime Video used cost-cutting techniques to reduce infrastructure and audio-video monitoring service costs by rearchitecting the system \cite{bib41}.

\subsubsection{Background}\label{sec:prime-bg}
The audio-video monitoring service at Prime Video was originally designed as a distributed microservices architecture. This architecture consisted of a number of independent services, each of which was responsible for monitoring a specific aspect of audio or video quality. For example, one service may be in charge of monitoring audio loudness, while another service may be in charge of monitoring video bitrate. The microservices architecture had several benefits. It was relatively simple to develop and deploy new services, and scaling the service by adding more instances of each service was simple. However, the microservices architecture had several drawbacks. It was difficult to manage the service because there were so many independent services to keep track of. Furthermore, the service was not very scalable because each service had to be scaled separately.

Therefore, Prime Video redesigned its audio-video monitoring service as a monolith to address these challenges. A monolith is a centralized service in charge of all aspects of audio or video quality monitoring. By consolidating all of the services into a single monolith, the audio-video monitoring service was made easier to scale and manage. This eliminated the need to duplicate data between different services, which made the service more efficient.

\subsubsection{Cost Optimization Techniques}\label{sec:prime-cost-opt}
AWS Step Functions is a serverless orchestration service for coordinating the execution of multiple AWS services. It is a powerful tool, but it is not cheap. Prime Video's audio-video monitoring service orchestrated the flow of data through the service using AWS Step Functions. This resulted in a major bottleneck issue.

Every second of the stream, the service went through multiple state transitions. As a result, account limits on Prime Video were quickly reached. Because AWS Step Functions charges users per state transition, the total cost of all the building blocks was too high for the solution to be adopted on a large scale.  Additionally, Amazon S3 was being used to store video frames by Prime Video's audio-video monitoring service and the service made a large number of Tier-1 requests to Amazon S3. Tier-1 calls are the most expensive type of Amazon S3 calls that can be made. As a result, Amazon S3 storage of video frames was costing Prime Video a lot of money. Furthermore, the number of video frames that must be stored can vary depending on traffic volume. As a consequence, scaling the number of Amazon S3 calls that the service could make was impossible.

Prime Video redesigned the architecture of its audio-video monitoring service to a monolith application to address bottlenecks and cost issues. Because all of the components were now running in a single process, AWS Step Functions and Amazon S3 were no longer required. As a result, Prime Video no longer had to pay for state transitions or Tier-1 calls, which resulted in significant cost savings. The monolith architecture also allowed the service to be scaled up and down as needed. This was critical because the service was expected to handle an increase in traffic. Finally, moving the solution to AWS EC2 and AWS ECS also enabled Prime Video to take advantage of AWS EC2's long-term compute savings plans as well as EC2 features like autoscaling based on the traffic load, which helped drive costs even lower.

\subsubsection{Results}\label{sec:prime-results}
Prime Video was able to reduce the cost of running its audio-video monitoring service by 90\% as a result of these changes. The service was also capable of handling significantly more traffic.

The following are some key takeaways from this case study of Amazon Prime Video:

\begin{itemize}
\item Architectures based on distributed microservices can be costly and difficult to scale.
\item Monolithic architectures have the potential to be more cost-effective and scalable.
\item Compute saving plans can assist you in saving money on Amazon EC2 usage.
\item Spot instances, reserved instances, and autoscaling can all help you cut your Amazon EC2 costs even further.
\end{itemize}

\subsection{Pinterest}\label{sec:pinterest}
Pinterest is a social media platform where users can share and discover ideas via images and videos. The company has been rapidly expanding in recent years, and as a result, its infrastructure costs have risen. This section explains how Pinterest used cost-cutting techniques to reduce infrastructure and stream processing costs in the cloud \cite{bib42}.

\subsubsection{Background}\label{sec:pinterest-bg}
Pinterest runs multiple Flink jobs in various sizes and importance across their production YARN clusters. These jobs do everything from compute engagement statistics to process long-tail data. However, managing these jobs in a multi-tenanted environment while ensuring efficiency, resource availability, and interference minimization presented significant challenges. Pinterest identified several key issues related to cluster configuration and resource utilization during their cost-cutting journey. The lack of CPU isolation was one of the major challenges, resulting in unstable load tests and CPU bursts from one job affecting others on the same host. Maladjusted VCore reservations and burst capacity allocation also had an impact on resource utilization and overall cost effectiveness. 

Pinterest undertook a series of initiatives to address these challenges and further optimize their Flink data processing clusters, including the implementation of CGroups soft CPU limits, capacity reservations, and container placement optimization. However, one of the most important steps they took was to switch from AWS i3 instances to i4i instances. Pinterest discovered that the new i4i instances performed exceptionally well with their Flink jobs, resulting in a 40\% reduction in CPU usage for a marginal 10\% cost increase. This upgrade not only improved performance, but it also reduced their overall platform's AWS spend by 10\%. By leveraging these strategies, including the adoption of i4i instances, Pinterest aimed to achieve better stability, improve resource utilization, and cost savings.

The following sections will delve into the specific steps Pinterest took to overcome these challenges and optimize their Flink data processing clusters, emphasizing the impact of their efforts on cost savings and performance improvements.

\subsubsection{Cost Optimization Techniques}\label{sec:pinterest-cost-opt}
Pinterest embarked on an attempt to reduce the cost of their Flink data processing clusters. They aimed to improve stability, improve resource utilization, and achieve significant cost savings through a variety of initiatives and strategies. This section delves into Pinterest's key techniques and their impact on cost optimization.

\paragraph{\textbf{Implementing CGroups Soft CPU Limits}}
Pinterest implemented CGroups soft CPU limits for each worker on their YARN clusters to address the lack of CPU isolation and mitigate noisy neighbor issues. They ensured burst capacity was available when needed by configuring soft limits rather than hard limits, particularly during job deployments and unexpected influxes of events. This approach enabled Pinterest to run the cluster at a higher capacity without sacrificing availability, resulting in a significant reduction in resource requirements and allowing them to downsize their clusters by 20\%.

\paragraph{\textbf{Hot Node Mitigation}}
Pinterest experienced problems with misaligned vcore reservations and hot nodes, which resulted in suboptimal resource utilization and CPU overcommitment. They reevaluated their CPU reservations for each job to ensure an appropriate balance between requested cores and actual usage. They effectively mitigated hot nodes and eliminated resource contention among jobs by enforcing CPU reservations and leveraging soft CPU limits. This not only improved stability but also reduced resource consumption, resulting in additional cost savings.

\paragraph{\textbf{Burst Capacity Policy Optimization}}
Pinterest improved their burst capacity policy by taking a more deliberate approach. Rather than reserving the entire burst capacity quota for each job, they aimed to efficiently allocate burst overhead based on actual burst needs. They ensured that burst resources were available to jobs during traffic peaks without triggering overload by implementing guaranteed shared burst capacity reservations and optimizing resource placement. This method improved cost efficiency by avoiding wasteful provisioning and allowed for better utilization of burst capacity.

\paragraph{\textbf{Data Processing Job Optimization}}
Pinterest were focused on making their Flink jobs faster and leaner. CPU banding was identified as a significant issue, resulting in inefficient resource utilization. They enforced a uniform distribution of tasks across Taskmanagers by streamlining task placements and leveraging colocation constraints, reducing CPU banding and improving overall CPU utilization. As a result, they were able to significantly reduce cross-host network traffic, CPU requirements, and per-job costs. These optimizations resulted in a 50-90\% cost reduction without sacrificing performance.

\paragraph{\textbf{Transition to AWS i4i Instances}}
Pinterest conducted extensive testing after recognizing the performance benefits of AWS i4i instances and concluded that these instances were highly efficient for their Flink jobs. The AWS i4i instances are the most recent generation of general-purpose EC2 instances built on the AWS Nitro System \cite{bib43}. They have several advantages over i3 instances \cite{bib44}, including:

\begin{itemize}
    \item \textbf{Better CPU performance}: AWS EC2 i4i instances are powered by third generation Intel Xeon Scalable processors, which provide up to 30\% better compute price than i3 instances, which use second generation Intel Xeon Scalable processors.
    \item \textbf{Lower cost per vCPU}: The i4i instances are priced at a lower cost per vCPU than i3 instances, making them a more cost-effective choice for many workloads.
    \item \textbf{Support for AWS Nitro SSD}: The i4i instances with AWS Nitro SSDs deliver up to 60\% lower storage I/O latency and up to 75\% reduced storage I/O latency variability than third generation EC2 storage optimized instances. This means that the applications will experience faster response times when accessing data from the disk.
\end{itemize}

By transitioning from i3 instances to i4i instances, they achieved a substantial 40\% reduction in CPU usage at a slightly increased cost of 10\%. This hardware upgrade proved to be a cost-effective decision, allowing Pinterest to optimize their AWS spend and boost their cost-cutting efforts.

\subsubsection{Results}\label{sec:pinterest-results}
Pinterest's efforts to optimize their Flink data processing clusters produced impressive results, including significant cost savings and performance improvements. They achieved better stability and resource utilization by implementing various techniques such as CGroups soft CPU limits, capacity reservation fixes, and container placement optimizations, resulting in a 20\% cost reduction. Furthermore, switching from AWS i3 instances to i4i instances resulted in a 40\% reduction in CPU usage. They achieved a 60\% reduction in cross-host network traffic and a 50\% reduction in CPU needs by mitigating CPU banding issues through improved task placements and leveraging colocation constraints. These measures, when combined with data optimization efforts, resulted in a 35\% cost reduction on the Stream Processing Platform. \\

\noindent The following are some key takeaways from this case study of Pinterest:
\begin{itemize}
    \item Adopt granular resource allocation to reduce waste and increase cost efficiency.
    \item Utilize hardware upgrades, such as switching to more efficient instance types, to save money.
    \item Solve multi-tenancy issues with techniques such as container placement optimization and burst capacity reservations.
    \item Improve resource utilization and reduce CPU banding by optimizing task placement and reducing CPU banding.
    \item Maintain long-term cost savings and performance by continuously evaluating and optimizing.
\end{itemize}

\subsection{Baselime: Cloud Platform Migration}\label{sec:baselime}
Baselime, an observability platform acquired by Cloudflare in 2024, provides a compelling case study of cost optimization through strategic cloud platform migration. This section examines how Baselime achieved over 80\% cost reduction by migrating from AWS to Cloudflare's developer platform \cite{bib54}.

\subsubsection{Background}\label{sec:baselime-bg}
Baselime originally built their observability platform entirely on AWS, utilizing services including AWS Lambda for data reception, Amazon Kinesis Data Streams for event streaming, Amazon CloudFront for content delivery, and self-hosted ClickHouse on EC2 instances for analytics. While this architecture was functional, the costs associated with I/O-bound Lambda functions and data streaming were substantial.

\subsubsection{Cost Optimization Techniques}\label{sec:baselime-cost-opt}
The migration strategy focused on replacing AWS services with Cloudflare equivalents that offered more favorable pricing models for their specific workload characteristics:

\begin{itemize}
    \item \textbf{Data Receptors Migration}: Baselime migrated their data reception layer from AWS Lambda to Cloudflare Workers. Since Workers charge based on CPU time rather than total execution duration, and the data receptors were primarily I/O-bound (moving data rather than processing it), this resulted in dramatic cost savings.
    \item \textbf{Analytics Engine Migration}: The self-hosted ClickHouse cluster on EC2 was replaced with Cloudflare's Workers Analytics Engine, eliminating the need for EC2 instances, disk storage, and Kinesis Data Streams.
    \item \textbf{CDN Elimination}: CloudFront costs were eliminated entirely as Cloudflare's network handled content delivery natively.
\end{itemize}

\subsubsection{Results}\label{sec:baselime-results}
The migration, completed in stages during mid-2024, achieved remarkable cost reductions:

\begin{itemize}
    \item \textbf{Data Receptors}: AWS Lambda costs reduced by over 85\%, from approximately \$790/day to an estimated \$25/day on Cloudflare Workers a 95\% reduction.
    \item \textbf{Analytics Infrastructure}: Combined EC2 and Kinesis costs reduced by over 95\%, from approximately \$1,150/day to an estimated \$300/day on Workers Analytics Engine a 70\% reduction.
    \item \textbf{Overall}: Total cloud costs reduced by over 80\%, while simultaneously improving query performance and enabling higher event throughput.
\end{itemize}

The key insight from this case study is that pricing model alignment matters significantly: workloads that are I/O-bound rather than CPU-bound can achieve dramatic savings by selecting platforms that charge based on CPU time rather than total execution duration.

\subsection{Netflix: Database Consolidation and Migration}\label{sec:netflix}
Netflix's consolidation of relational database infrastructure on Amazon Aurora represents a strategic approach to reducing operational complexity while improving performance and cost efficiency. This case study examines how Netflix achieved up to 75\% performance improvements and 28\% cost savings through database migration \cite{bib55}.

\subsubsection{Background}\label{sec:netflix-bg}
Netflix's Online Data Stores (ODS) team faced significant challenges with their fragmented relational database strategy. Managing multiple PostgreSQL-compatible engines, including a licensed self-managed distributed PostgreSQL-compatible database as their primary solution, created operational inefficiencies that impacted both infrastructure teams and developers. The infrastructure team was burdened with self-managed databases on Amazon EC2, consuming valuable time with operational overhead from deployments, patching, scaling, and maintenance activities while facing rising licensing costs. The developer experience suffered from inconsistent database deployment processes across multiple engines, manual scaling procedures during traffic spikes, and the need to maintain expertise across multiple systems.

\subsubsection{Cost Optimization Techniques}\label{sec:netflix-cost-opt}
Netflix's migration to Amazon Aurora PostgreSQL addressed these challenges through several strategic approaches. The team established evaluation criteria across four key dimensions: developer productivity (PostgreSQL compatibility, minimal code changes), operational efficiency (simplified replica management, full infrastructure abstraction), performance reliability (high availability, automatic storage scaling, multi-Region reader support), and cost efficiency (lower total cost of ownership, ability to support expanding workloads). Aurora's shared storage architecture eliminated the cross-Availability Zone latency overhead present in their previous distributed solution, allowing the database engine to allocate 75\% of instance memory to shared buffers compared to the typical 25-40\% in standard PostgreSQL. The pay-as-you-go pricing model, combined with features like storage auto-scaling up to 256 TB and continuous incremental backup to Amazon S3, removed manual capacity management requirements.

\subsubsection{Results}\label{sec:netflix-results}
As of October 2025, Netflix has migrated several applications from their self-managed distributed PostgreSQL-compatible database to Aurora PostgreSQL, achieving substantial improvements. For Spinnaker's Front50 metadata microservice, the migration delivered approximately 50\% reduction in average latency (from 67.57 milliseconds to 41.70 milliseconds), approximately 70\% reduction in maximum latency with fewer spikes, and much more consistent performance patterns. The Policy Engine, Netflix's rules engine for data governance, saw even more dramatic improvements: countDatasets latency reduced from 5.40 milliseconds to 1.90 milliseconds, findDatasets from 26.72 milliseconds to 6.51 milliseconds, and getAggregatedFilterTerms from 12.11 milliseconds to 3.51 milliseconds. Overall, Netflix achieved up to 75\% performance improvements and 28\% cost savings through Aurora's pay-as-you-go pricing model compared to license-based pricing, while eliminating significant operational overhead.

\subsection{Comparative Analysis}\label{sec:case-study-analysis}
The four case studies Prime Video, Pinterest, Baselime, and Netflix highlight diverse optimization strategies across different scales and contexts. While Prime Video and Pinterest focus on optimizing existing cloud deployments through architectural changes and resource tuning, Baselime and Netflix demonstrate the potential of strategic platform selection and migration. Here are some points of comparison:

\subsubsection{Scope}: Pinterest's cost-cutting efforts are focused primarily on Flink job management and resource allocation within their YARN clusters. They must balance the overall system's efficiency, ensure resource availability for higher-tier jobs, and prevent job interference in a multi-tenant environment. This includes improving CPU reservations for jobs of varying importance and scale, as well as optimizing CPU utilization and addressing noisy neighbor issues. On the other hand, Prime Video focuses on optimizing their video quality analysis system and monitoring infrastructure. Their goal is to reduce infrastructure costs while increasing the capacity of their defect detection system to handle thousands of concurrent streams. They address issues such as scaling bottlenecks, high costs of distributed components, and orchestration management limitations.

\subsubsection{Optimization Targets}: The optimization efforts of Prime Video are aimed at lowering infrastructure costs and improving scaling capabilities. They intend to seamlessly monitor thousands of streams while minimizing the costs associated with distributed components and orchestration management. The goal is to handle increasing loads efficiently while also providing their customers with a high-quality streaming experience. On the contrary, Pinterest's optimization efforts are aimed at improving the overall efficiency of their system. They intend to ensure that higher-tier jobs have the resources they require, to prevent job interference in the multi-tenant YARN environment, and to improve the overall efficiency of their Flink jobs.

\subsubsection{Architectural Changes}: Pinterest's multi-tenant YARN clusters address a lack of CPU isolation and inefficient capacity reservations. They use CGroups soft CPU limits to enforce CPU limits proportional to requested capacity, ensuring that burst capacity is regulated. They improve multi-tenant stability and prevent noisy neighbor issues by leveraging CPU-aware scheduling and introducing guaranteed burst capacity reservations. Prime Video redesigns their system to move away from a distributed microservices approach and toward a monolithic application. They eliminate the need for intermediate storage and reduce data transfer costs by combining all components into a single process. They use a single instance to implement an orchestration layer, which improves scalability and simplifies control flow.

\subsubsection{Resource Utilization}: Both case studies emphasize the importance of optimizing resource utilization. Prime Video achieves resource utilization improvements by consolidating their components into a monolithic application. They eliminate the need for expensive video frame storage and reduce computational overhead by transferring data within memory. This enables them to process and analyze streams more efficiently, resulting in cost savings and improved performance. Pinterest prioritizes resource utilization in their Flink jobs. They address CPU banding issues by optimizing task placement, reducing cross-host network traffic, and enforcing colocation constraints. This results in more balanced CPU utilization and efficient resource utilization, resulting in cost savings without sacrificing performance.

\subsubsection{Cost Reduction}: Both companies achieve significant cost savings through their optimization efforts. Pinterest's optimization efforts result in significant cost savings. They are able to reduce their cluster size by 20\% by implementing CGroups soft CPU limits and optimizing container placement. Furthermore, hardware upgrades to AWS i4i instances increase the efficiency of Flink jobs by 40\%, resulting in cost savings and better resource utilization.
By switching to a monolithic architecture, Prime Video achieves significant cost savings. The consolidation of components and the reduction of data transfer costs result in infrastructure cost savings of more than 90\%. Using Amazon EC2 and Amazon ECS instances optimizes their cost structure even further, and they can benefit from Amazon EC2 compute saving plans for even more cost savings.

Both Pinterest and Prime Video demonstrate the effectiveness of their cost optimization strategies in their respective systems by focusing on these optimization targets, implementing architectural changes, improving resource utilization, and achieving significant cost reductions.

\subsection{Key Takeaways}\label{sec:case-study-key-takeaway}
The following key takeaways can be drawn from the four case studies Prime Video, Pinterest, Baselime, and Netflix:

\begin{itemize}
    \item \textbf{Efficient Resource Allocation and Utilization}: Both case studies emphasize the importance of optimizing resource allocation and utilization in order to save money and improve system performance. Pinterest and Prime Video both show the value of fine-tuning resource allocation, such as CPU utilization and resource placement, in order to eliminate waste and improve efficiency.
    
    \item \textbf{Considerations for System Architecture}: Reevaluating system architecture can help to address scalability issues and cut costs. Pinterest's adoption of CGroups soft CPU limits and optimized resource placement, as well as Prime Video's transition to a monolithic application, demonstrate the importance of architectural changes in achieving scalability, cost reduction, and orchestration logic simplification.
    
    \item \textbf{Technology Stack Evaluation}: It is critical to select the appropriate technology stack for the specific use case. Organizations should weigh the benefits and drawbacks of various architectural approaches, taking into account factors such as scalability, cost, and ease of management. The case studies of Pinterest and Prime Video highlight the importance of selecting technologies that align with optimization goals and support efficient resource utilization.
    
    \item \textbf{Fine-tuning and Optimization}: Continuous fine-tuning and optimization of job parameters, task placements, and algorithmic approaches can result in significant cost savings while maintaining performance. The impact of such optimizations can be seen in Pinterest's efforts to eliminate CPU banding and reduce cross-host network traffic, as well as Prime Video's focus on reducing data transfer and computational overhead.
    
    \item \textbf{Continuous Improvement}: Cost optimization is a continuous process that necessitates ongoing monitoring, experimentation, and adaptation. Organizations should create feedback loops, analyze system performance, and proactively identify opportunities for optimization. Organizations can achieve long-term cost efficiency and stay ahead of changing requirements by continuously improving their systems.
    
    \item \textbf{Balancing Cost and Quality}: While cost reduction is important, it is also important to maintain or improve service quality and customer experience. Along with cost optimization, all four case studies prioritize providing a seamless and high-quality user experience. Long-term success requires striking the right balance between cost and quality.
    
    \item \textbf{Platform and Pricing Model Alignment}: Baselime's migration demonstrates that selecting platforms with pricing models aligned to workload characteristics can yield dramatic savings. I/O-bound workloads benefit significantly from platforms that charge based on CPU time rather than total execution duration.
    
    \item \textbf{Managed Service Migration}: Netflix's database consolidation demonstrates that migrating from self-managed infrastructure to managed services can deliver both cost savings and operational improvements. Leveraging fully managed database services with features like automatic storage scaling, continuous backup, and infrastructure abstraction enables organizations to reduce operational overhead while improving performance and cost efficiency.
\end{itemize}

By considering these key takeaways, organizations can learn from the experiences of Prime Video, Pinterest, Baselime, and Netflix, and apply similar strategies to optimize their systems for cost efficiency while ensuring high performance and quality. A proactive and iterative cost optimization approach, combined with a thorough understanding of system requirements and workload characteristics, can result in significant cost savings and improved overall efficiency.

\section{Future Research Considerations}\label{sec:future-work}
The cost optimization techniques as well as case studies discussed in this paper offer valuable insights into cost optimization strategies and architectural improvements. The rapid evolution of AI infrastructure has introduced new research challenges and opportunities. These discoveries lay the groundwork for future research in the fields of cost optimization and system scalability. Here are some possible areas for further investigation:

\subsubsection{Automated System Monitoring and Optimization}
One area of focus for the future research could be the creation of automated monitoring and optimization frameworks. Future research could look into developing frameworks that continuously analyze system performance, identify inefficiencies, and recommend changes to resource allocation and job configurations. Using artificial intelligence and machine learning techniques, the optimization process can be automated, allowing systems to adapt and optimize in real-time. Recent work on ABACUS (Automated Budget Analysis and Cloud Usage Surveillance) demonstrates the potential of automated FinOps services that enforce budgets, alert teams of spending breaches, and leverage Infrastructure-as-Code to predict deployment costs before resources are provisioned \cite{bib57}.

\subsubsection{Advanced Resource Allocation Techniques}
Future research could focus on advanced resource allocation techniques that go beyond traditional methods. Exploring machine learning algorithms or optimization models to dynamically allocate resources based on workload characteristics, job priorities, and cost constraints could be part of this. To optimize resource utilization and reduce costs, techniques such as predictive resource allocation and proactive resource provisioning can be investigated.

\subsubsection{Cost-Performance Trade-off Analysis}
Further research can be carried out to conduct in-depth studies on the trade-off between cost optimization and system performance. It is critical to assess the impact of various cost-cutting measures on overall system performance, user experience, and quality. Researchers can develop methodologies and models that strike a balance between cost reduction and maintaining or improving service quality by understanding the complexities of this trade-off.

\subsubsection{Adaptive Scaling and Bursting}
Adaptive scaling techniques, which allow systems to dynamically adjust resource allocation based on real-time workload demands, could be the focus of future research. Exploring strategies for efficient bursting during peak times and scaling down during off-peak times can help to maximize cost utilization. Investigating auto-scaling algorithms that take workload patterns, historical data, and cost constraints into account can lead to more efficient resource allocation.

\subsubsection{Multi-Cloud and Hybrid Cloud Cost Optimization}
As organizations increasingly adopt multi-cloud and hybrid cloud environments, future research can look into cost-cutting strategies tailored to these configurations. Investigating techniques for dynamically scaling resources across multiple cloud providers or combining on-premises infrastructure with cloud resources can aid in cost optimization. It will be beneficial to investigate cost optimization models that take into account the unique characteristics of multi-cloud and hybrid cloud architectures.

\subsubsection{Cost Modeling and Predictive Analytics}
Cost modeling techniques and predictive analytics models can help forecast resource usage and estimate the impact of various optimization strategies on cost savings. Methods for accurately modeling and forecasting resource demands, cost trends, and utilization patterns can help organizations make informed decisions and plan their resource allocation strategies more effectively.

\subsubsection{AI Infrastructure Cost Optimization}
The rapid growth of AI workloads presents unique cost optimization challenges that warrant dedicated research attention. Key areas include:
\begin{itemize}
    \item \textbf{Intelligent model routing}: Developing systems that automatically route inference requests to the most cost-effective model based on query complexity and quality requirements.
    \item \textbf{Dynamic quantization}: Research into adaptive quantization techniques that adjust model precision based on workload characteristics and cost constraints.
    \item \textbf{GPU sharing and multiplexing}: Investigating techniques for efficiently sharing GPU resources across multiple workloads to improve utilization.
    \item \textbf{Inference cost prediction}: Building models that accurately predict inference costs for complex AI pipelines to enable better capacity planning.
    \item \textbf{Training efficiency}: Exploring techniques such as curriculum learning, data pruning, and efficient architectures to reduce training costs.
\end{itemize}

\subsubsection{LLM-Specific Cost Optimization}
As large language models become ubiquitous in enterprise applications, research into LLM-specific cost optimization is increasingly important:
\begin{itemize}
    \item \textbf{Prompt optimization}: Techniques for automatically compressing or optimizing prompts to reduce token usage while maintaining output quality.
    \item \textbf{Speculative decoding}: Research into speculative decoding and other techniques that can reduce inference latency and cost.
    \item \textbf{Model distillation}: Developing efficient distillation techniques to create smaller, cheaper models that maintain the capabilities of larger models for specific use cases.
    \item \textbf{Semantic caching}: Advanced caching strategies that leverage semantic similarity to reduce redundant inference calls.
\end{itemize}

\subsubsection{Sustainability and Green Computing}
The environmental impact of cost-cutting strategies is also an important area of investigation. The growth of AI workloads has significantly increased data center energy consumption, with McKinsey estimating that AI data center infrastructure could require \$7 trillion in investment through 2030. Exploring methods to reduce energy consumption and carbon footprint can help to ensure the long-term viability of cloud computing. Green computing and cost optimization can be advanced by researching energy-efficient resource allocation algorithms, investigating techniques for dynamic power management, and investigating the integration of renewable energy sources for powering cloud infrastructure.

\subsubsection{FinOps Maturity and Automation}
As FinOps practices mature, research into advanced automation and governance frameworks becomes essential. Areas of investigation include automated policy enforcement, anomaly detection for cost spikes, and integration of cost optimization into CI/CD pipelines. The emergence of FinOps for AI as a distinct discipline also presents opportunities for research into specialized tools and methodologies for managing AI infrastructure costs.

Organizations and researchers can advance the field of cost optimization, scalability, and efficiency in cloud computing systems by delving into these future research considerations. Continued research and development in these areas will lead to more sustainable, cost-effective, and high-performing systems in the future.

\section{Conclusion}\label{sec:conclusion}
Understanding and effectively navigating cloud and AI infrastructure pricing models is essential for organizations seeking to maximize the value of their technology investments. This paper has provided a comprehensive review of cost optimization strategies spanning traditional cloud computing and the rapidly evolving AI infrastructure landscape.

Case studies from Prime Video, Pinterest, Baselime, and Netflix demonstrate that organizations can achieve 28-90\% cost reductions through strategic architectural decisions, platform selection, database consolidation, and pricing model alignment. The emergence of AI workloads presents unique challenges, with GPU compute representing 40-60\% of technical budgets, yet the economics are improving rapidly LLM inference costs have decreased by approximately 10x annually since 2021. Organizations can leverage techniques such as model quantization, intelligent routing, and efficient fine-tuning to significantly reduce AI infrastructure costs.

The growth of FinOps practices, with 59\% of organizations now maintaining dedicated teams, reflects the increasing importance of cloud financial management. By combining pricing model selection with robust cost management strategies, organizations can achieve cost efficiency while maximizing the potential of cloud and AI infrastructure.
\bibliographystyle{splncs03_unsrt}
\bibliography{references}

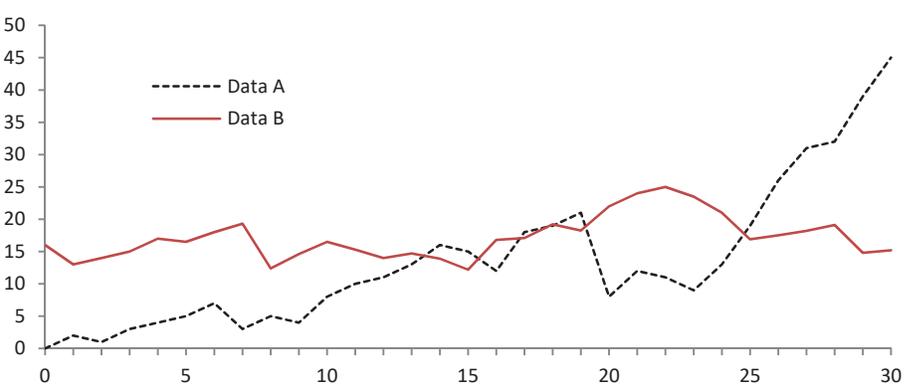
\end{document}